\documentclass{ws-mpla}

\newcommand{\stau}{\mbox{$\tilde{\tau}$}}
\newcommand{\neutralino}{\mbox{$\chi^0_1$}}
\newcommand{\chargino}{\mbox{$\chi^\pm_1$}}

\begin{document}

\markboth{T. Adams}
{SEARCHES FOR LONG-LIVED PARTICLES AT THE TEVATRON COLLIDER}

\catchline{}{}{}{}{}

\title{SEARCHES FOR LONG-LIVED PARTICLES AT THE TEVATRON COLLIDER}

\author{\footnotesize TODD ADAMS}

\address{Department of Physics, Florida State University, 515 Keen Building\\
Tallahassee, Florida 32306,
U.S.A. \\ 
tadams@hep.fsu.edu}



\maketitle

\pub{Received (Day Month Year)}{Revised (Day Month Year)}

\begin{abstract}
Several searches for long-lived particles have been performed
using data from $p\bar{p}$ collisions from Run II at the 
Tevatron.  In most cases, new analysis techniques have
been developed to carry out each search and/or estimate
the backgrounds. These searches expand the discovery 
potential of the CDF and D0 experiments to new physics that 
may have been missed by traditional search techniques.  This
review discusses searches for (1) neutral, long-lived particles
decaying to muons, (2) massive, neutral, long-lived particles
decaying to a photon and missing energy, (3) stopped 
gluinos, and (4) charged massive stable particles.  It
summarizes some of the theoretical and experimental motivations
for such searches.

\keywords{searches; long-lived; supersymmetry.}
\end{abstract}

\ccode{PACS Nos.: 13.85.Rm, 14.80.Ly, 14.80.Cp.}

\section{Introduction}	

The discovery of new particle phenomena has led to 
some of the greatest advances in our
understanding of particle physics, particularly when
the discovery is unexpected.  Among the most famous 
are the muon (1937), strange mesons (``V'' particles,
1947), the $J/\psi$ meson (1974), and the tau lepton (1974).  
There is broad agreement among particle physicists that
progress in understanding some fundamental questions,
such as the origin of mass, requires new physics.
While there are many interesting theories of possible
new physics, a few of which we shall touch upon
in this review, no one actually knows what Nature 
has in store for us.  Therefore, it is
imperative that modern particle physics experiments
search for signs of new physics in as many different ways as
possible, including those that are motivated
by current theory and those that are not.

Numerous searches for new particle phenomena have been
done using data from the Tevatron, the Fermi National 
Accelerator Laboratory (Fermilab) hadron collider.
Most of these searches focus on reconstructing particles
that decay within millimeters of the beamline or on 
observing missing transverse energy signatures from stable 
particles that escape without interacting.  However, as
is shown in this review, the multipurpose design of the 
Tevatron detectors, D0 and CDF, allows a broader range
of searches to be conducted, namely for
long-lived, but unstable particles.  Many possibilities
exist that give rise to unique signatures.

\subsection{Theoretical Motivation \label{sec:theoryintro}}	

A wide variety of theoretical models, including large
classes based on supersymmetry (SUSY), allow for long-lived
particles.  The simplest SUSY variation is 
$R$-parity violation which allows the lightest
supersymmetric particle (LSP) to decay purely to 
non-SUSY particles.\cite{bib:rpvsusy}  
The rate of such decays is generally determined by
one or more couplings.  Numerous searches at both the
LEP and Tevatron colliders have looked for (nearly)
prompt $R$-parity violating processes, however they are
insensitive to longer lifetimes where the coupling
is small.\cite{bib:rpvsearches1,bib:rpvsearches2,bib:rpvsearches3,bib:rpvsearches4,bib:rpvsearches5,bib:rpvsearches6} 

Gauge mediated supersymmetry breaking (GMSB) models,
in which the supersymmetry breaking is mediated by gauge
fields other than gravity,
generally have a light (MeV) gravitino LSP and all
other supersymmetric particles decay via cascades
to the next-to-lightest supersymmetric 
particle (NLSP).  The coupling to gravity can 
determine how quickly the NLSP decays to its
non-SUSY partner and the gravitino.  If the
symmetry breaking scale is small (larger coupling),
the lifetime is short.  Conversely, if the scale is 
large, this results in a long-lived NLSP that can decay 
either within the detector or beyond it.\cite{bib:feng}  
Favored NLSP candidates include the 
neutralino (\neutralino) or the stau (\stau).

Long lifetimes can also arise from small mass 
splittings between the NLSP and the 
LSP.\cite{bib:masssplitting}  If the mass
difference ($M_{NLSP} - M_{LSP}$) is small
enough, phase space suppression
will impede the decay of the NLSP.  One
particular model of interest here is inspired by
anomaly mediated supersymmetry breaking (AMSB)
with a chargino NLSP and a mass
splitting $<$ 150 MeV.

Split supersymmetry, where SUSY scalars are much heavier
than SUSY fermions, can suppress the decay rate of gluinos 
giving them long lifetimes.\cite{bib:splitsusy1,bib:splitsusy2}  
They may live long enough to hadronize into ``R-hadrons," 
colorless, bound states 
of the gluino, quarks, and gluons.\cite{bib:rhadrons}
Nuclear interactions can cause some neutral R-hadrons to become
charged, leading to a number of interesting signatures
within the detector.\cite{bib:howie}  

There has been recent discussions of a class of models
referred to as ``hidden valley'' models which can yield
new phenomena resulting in long-lived particles through 
reduced coupling to standard model (SM) 
particles.\cite{bib:strassler1,bib:strassler2,bib:strassler3}
The hidden valley structures are very general and can be
added to many beyond the standard model (BSM) theories.
Most interestingly, they can give rise to significantly 
different phenomenology from that commonly associated
with the original BSM model.\cite{bib:strassler1}
An example is a light Higgs boson that decays to two 
neutral, long-lived particles that further decay to 
$b$-jets.\cite{bib:strassler2}  
These ``hidden valley'' models provide a timely 
cautionary tale: if one
only looks where the light shines brightest, one risks
missing a possible spectacular prize in the shadows.

\subsection{Experimental Motivation \label{sec:nutev}}	

The NuTeV experiment at Fermilab observed an unexpected 
excess of events in one of three searches for long-lived
neutral particles.\cite{bib:nutev3events}  
The NuTeV neutrino detector was 
augmented with a 30 m long, low-mass decay region in front.
The decay region was instrumented with several drift
chambers along its length.  A weakly-interacting, long-lived
neutral particle produced by proton-nucleon interactions
from the 800 GeV Tevatron beam could travel 1.4 km and
decay in the low mass region.

NuTeV performed three separate searches (in mass bins)
by reconstructing a vertex from a pair of tracks from
hits in the decay region and using particle 
identification from the neutrino detector.
In the high mass (2.2-15 GeV) search, three events
were reconstructed with a vertex from
two muons and an unobserved neutrino.  The 
largest backgrounds originate from neutrino scattering
within the decay region and in the upstream shielding.  The
total estimated background was 0.07 $\pm$ 0.01 events,
which is to be compared with three observed events.

No widely accepted explanation of this excess has been
found.  
No backgrounds are large enough to explain it.
Cross-checks performed with events of similar topology 
within the neutrino detector support the hypothesis that
the events are unlikely to be due to neutrino scattering.  
If a new particle is responsible, the three events are 
consistent with a particle of $\approx$5 GeV mass 
decaying to $\mu^+\mu^-\nu$.
The best theoretical hypothesis for a new particle is a 
long-lived neutralino from $B$ mesons produced at the
primary proton target.\cite{bib:dedes}  Unfortunately, no current 
experiment matches NuTeV's running conditions so it has
been impossible to repeat the same search.  However,
as detailed below, collider experiments could, in
principle, observe such a
hypothetical particle, depending on the production
mechanism.

\subsection{The CDF and D0 Detectors}	

\begin{figure}[tbp]
 \centerline{\psfig{file=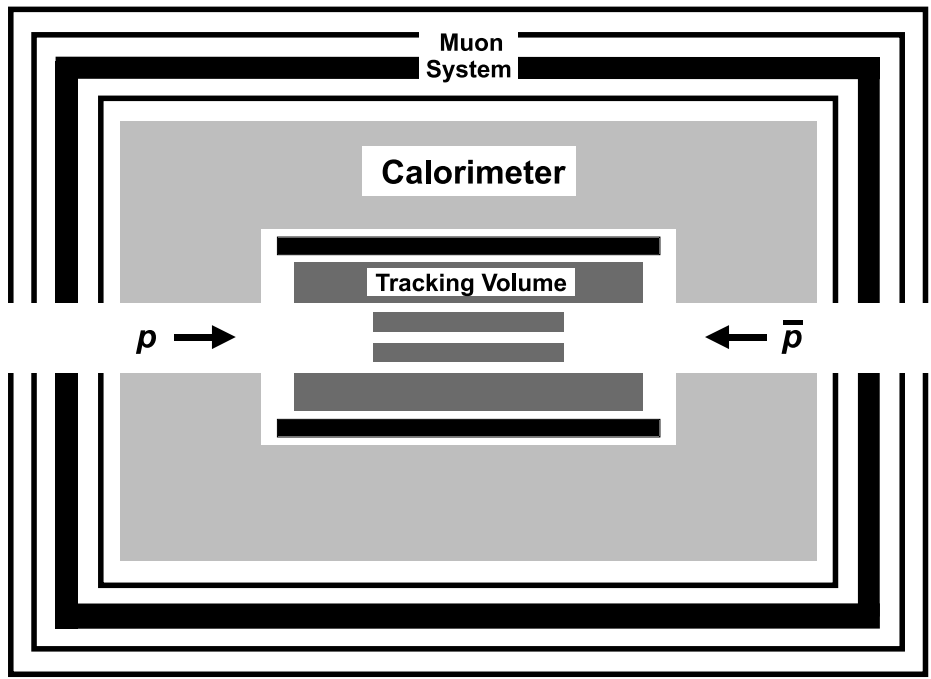,width=13cm}}
 \vspace*{8pt}
 \caption{A general systematic (side view) of the Tevatron collider detectors.
   The interaction region is at the center.  The gray
   indicates tracking volumes, the black boxes show solenoid (inner)
   and toroid (outer) magnets, the light gray region
   represents the calorimeter, and the black lines show the
   muon system.
 \protect\label{fig:detector}}
\end{figure}

While the specific detectors differ in technology and
implementation, the CDF\cite{bib:cdfdetector}
and D0\cite{bib:d0nim} detectors share similar 
general characteristics.  
The overall structure (from inside 
outward) consists of a central tracking region, a calorimeter,
and a muon system (see Fig.~\ref{fig:detector} for a
generic schematic).  The 
central tracking system consists of a silicon tracking detector
surrounded by another large volume tracker within a 
solenoid magnet.  The silicon tracker provides precision
vertex measurement as well as identification of heavy quarks
using secondary vertices.  The outer tracker, with its large
lever arm, measures the transverse momenta of charged
tracks.  Table~\ref{tab:detectors} lists some of the
specific characteristics of the individual detectors. 

\begin{table}
 \tbl{Summary of CDF and D0 detector technologies and
          characteristics. \label{tab:detectors}}
  {\begin{tabular}{lcc}
                           &        CDF         &         D0        \\ \toprule
    Inner Tracker          &      silicon       &      silicon      \\
    Outer Tracker          &    drift chamber   &   fiber tracker   \\
    Tracking Volume Radius & 1.3 $< r <$ 132 cm & 1.5 $< r <$ 50 cm \\
    Solenoid Magnet        &       1.4 T        &         2 T       \\
    Calorimeter            & iron/scintillator  & liquid-Ar/Uranium \\
    Muon System            & drift tubes and scintillator & drift tubes and scintillator \\
    Muon Toroid Magnet     &       1.4 T         &        1.8 T      \\
     \botrule
  \end{tabular}}
\end{table}

Collider detectors are optimized to reconstruct particles
that decay within millimeters of the beamline (including  bottom or
charm quarks) or to detect stable particles that escape without 
interacting (e.g. neutrinos) through the measurement of
missing transverse energy. However, they can also be used to 
search for particles that decay with long lifetimes.  Below
we review several analyses that use techniques including
detached vertex reconstruction, timing, and unique
signatures to search for such particles.

\section{Neutral Long-lived Particles Decaying to Two Muons}	

D0 has carried out a search for neutral particles that have
a (relatively) long lifetime and decay to two muons plus
missing transverse energy.\cite{bib:nllp_adams}  
One of the motivations for this search was to explore what
the Tevatron collider can say about the NuTeV result
described in Sec.~\ref{sec:nutev}.
The signature was a highly displaced vertex reconstructed 
using tracks in the central fiber 
tracker (CFT), which are matched to signals in the 
muon system.  The neutral particle must 
travel at least 5-20 cm in the transverse plane before
decaying.  These decay lengths are significantly longer
than those of normal $b$-hadrons.  Studies of $K_S \rightarrow
\pi^+\pi^-$ decays demonstrated the 
ability to find these highly displaced vertices.

Events with two isolated muons with $p_T >$ 10 GeV were
selected.  The muons were required to have a
distance of closest approach (DCA) to the primary
interaction vertex (PV) of greater than 0.01(0.1) cm
in the x-y plane (along the $z$ axis).  This reduces
events where the muons originate from the PV.
The muon tracks were fit to a common vertex and its
distance in the transverse dimension (i.e. the radius) 
was calculated from
$r = \sqrt{(X - X_{PV})^2 + (Y - Y_{PV})^2}$ where
$X_{PV},Y_{PV}$ are the $x$ and $y$ coordinates of the
primary vertex.  The final event selection requires
$5 < r < 20$ cm.

There are no SM sources of highly displaced vertices 
involving muon pairs, so the largest backgrounds originate
from mis-reconstruction.  Such instrumental backgrounds
are typically difficult to model accurately, therefore,
D0 estimated the background 
from data that closely (but not exactly) resemble 
those in the final selection.  Events 
with $0.3 < r < 5$ cm and/or only one muon passing
the DCA requirement were used.  The total
background was estimated to be $0.75 \pm 1.5$ events.

\begin{figure}[tb]
 \begin{center}
  \unitlength1cm
  \begin{picture}(12.5,4)(0,0)
   \put(0.0,0.0){\psfig{file=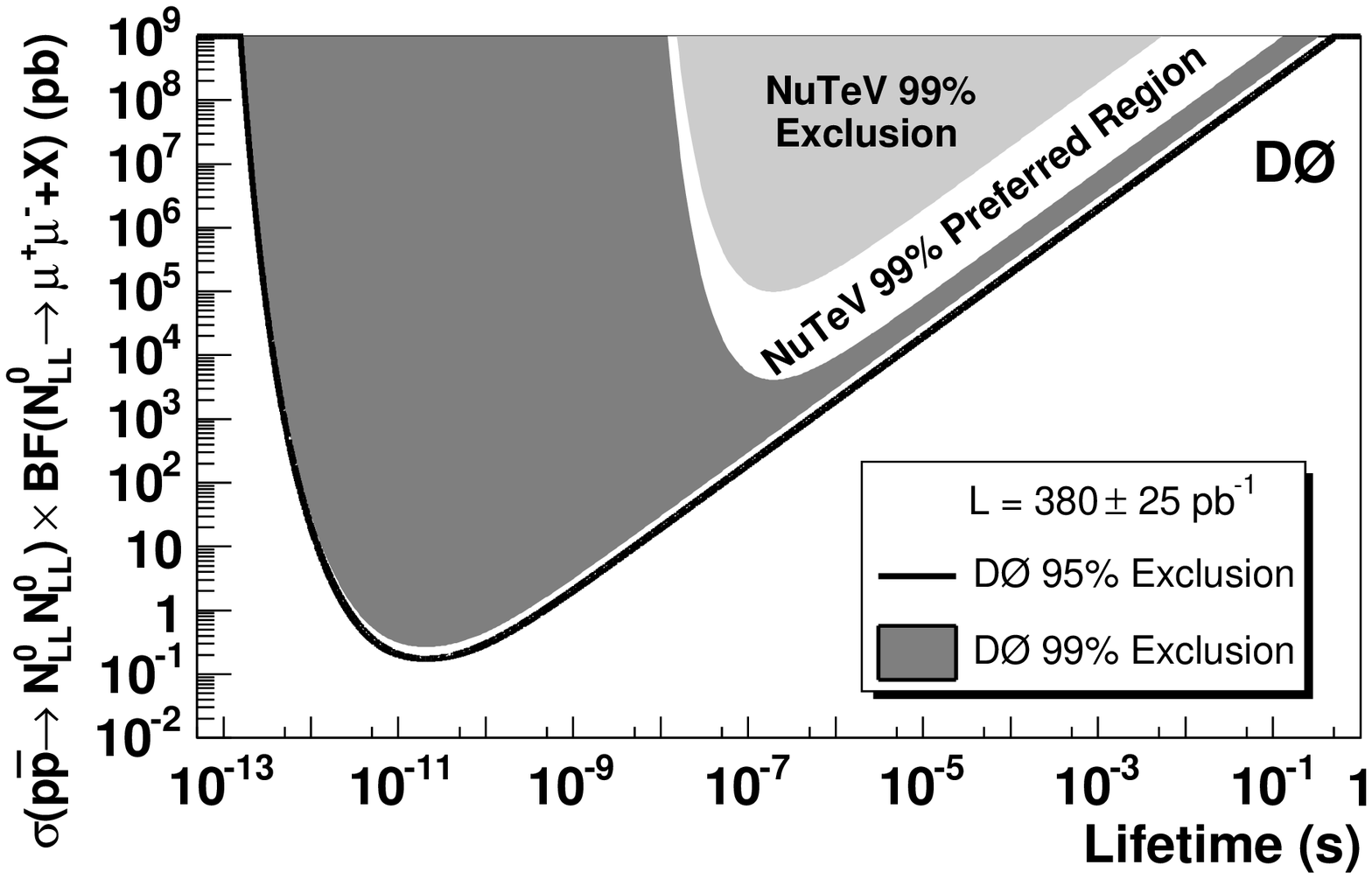,width=6cm}}
   \put(6.25,0.0){\psfig{file=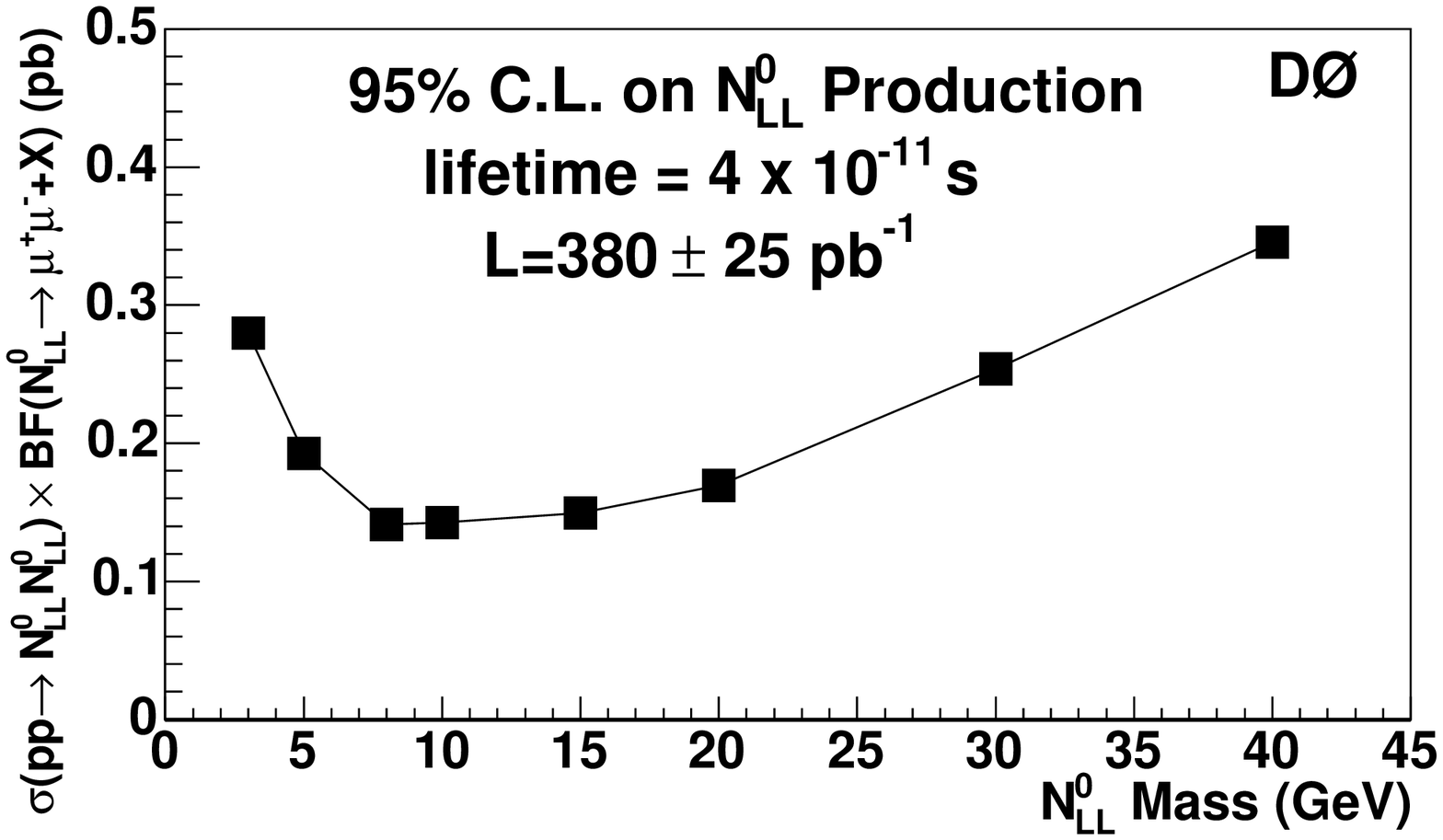,width=6cm}}
  \end{picture}
 \end{center}
\vspace*{8pt}
\caption{(left) The limit on the pair production cross section
 times branching ratio for a 5 GeV neutral, weakly-interacting
 particle that decays to two muons and an unobserved particle as
 a function of the particle lifetime.  The published NuTeV
 limit is converted to the D0 running
 conditions using an uMSSM production model.  The light band
 indicates the region favored by an interpretation of the three
 observed events as signal.  (right)  The limit as a function
 of particle mass for a lifetime of $4 \times 10^{-11}$ seconds.
 Reprinted figures with permission from V.~M.~Abazov {\it et al.}  
 [D0 Collaboration], Phys.\ Rev.\ Lett.\  {\bf 97}, 161802 (2006).
 Copyright 2006 by the American Physical Society.
 \protect\label{fig:d0nllplimits} }
\end{figure}

No events were found to pass all of the selection
criteria in data, therefore, limits on the production cross section
times branching ratios were set (Fig.~\ref{fig:d0nllplimits}).  
The limit depends on the lifetime and the mass (through
the acceptance).  A comparison was made with limits
set by NuTeV as well as with what one would expect if
the NuTeV excess is interpreted
as a signal.  To do so, the NuTeV limit was converted 
from that of $pp$ interactions
at $\sqrt{s}$ = 38 GeV to one for $p\bar{p}$ interactions at
$\sqrt{s}$ = 1960 GeV using a model of neutralino pair
production in an unconstrained minimal supersymmetric
model (uMSSM).\cite{bib:luibo}  As a result,
D0 can exclude some interpretations of the
NuTeV result at all lifetimes.

\section{Neutral Long-lived Particles Decaying to Photons}	

CDF performed a search for a heavy, long-lived particle
decaying to a photon and an unobserved particle.\cite{bib:cdfllphotons}
A heavy particle produced at the Tevatron would move slowly
away from the interaction point.  If the particle is
unstable, yet long-lived, its decay products would be 
delayed in reaching the various detectors relative to those
from prompt decays.\cite{bib:gunion}  CDF implemented measurement 
of the timing of particles arriving at its electromagnetic 
calorimeter with a timing resolution of 
0.6 ns.\cite{bib:cdfemtiming,bib:toback2004}  

The timing measurement formed the basis of a search
for delayed photons from decays of massive particles.  While
the search is model independent, the prototypical GMSB signal 
$\chi_1^0 \rightarrow \gamma G$, with a $\chi_1^0$ lifetime
of 5 ns was used to estimate a typical acceptance for such
events.  Events
were selected with at least one high $p_T$ ($>$30 GeV)
photon, one jet (which could also be from a photon), and
large missing transverse energy ($>$30 GeV).

The corrected arrival time of the photon ($t_c^{\gamma}$)
is defined as
$t_c^{\gamma} = t_{f} - t_{i} - \frac{|\vec{x}_{f} - \vec{x}_{i}|}{c}$
where $t_{i,f}$ and $\vec{x}_{i,f}$ are the initial 
and final photon time and position.  The initial 
position is given by the primary vertex while the 
final position is the arrival at the calorimeter.
Photons produced directly or via prompt decays have a 
Gaussian time distribution with a mean of zero.  The analysis
searches for an excess of events with 2 $< t_c^{\gamma} <$ 10 ns.

\begin{figure}[tbh]
 \begin{center}
   {\psfig{file=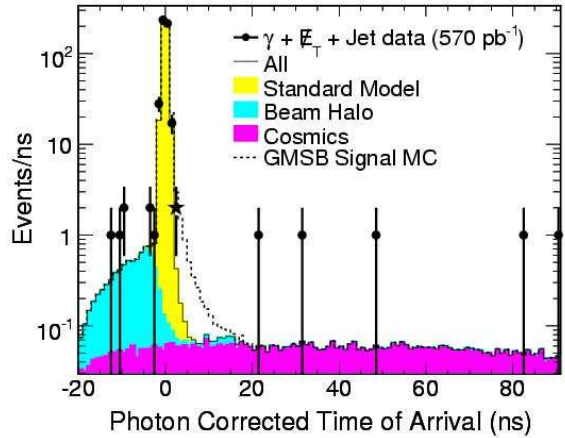,height=6.0cm}}
 \end{center}
\vspace*{8pt}
\caption{The corrected arrival time of the photon ($t_c^{\gamma}$)
 for data (black points), SM backgrounds (yellow),
 beam halo (blue), and cosmics (magenta).  Sample GMSB
 signal is shown as the dashed line.  The signal region
 corresponds to $t_c^{\gamma}$ = 2-10 ns.
 Reprinted figure with permission from A.~Abulencia {\it et al.}  
 [CDF Collaboration], Phys.\ Rev.\ Lett.\ {\bf 99}, 121801 (2007).
 Copyright 2007 by the American Physical Society.
 \protect\label{fig:cdfgammatime}}
\end{figure}

Standard model backgrounds from collisions include 
$\gamma$ + jet, dijet, and $W \rightarrow e\nu$ events.  In the
first two cases, the missing transverse energy must arise from energy
mis-measurement.  In dijet ($W$) events, a jet (electron) 
is mis-identified as a photon.
Non-SM backgrounds include cosmic ray muons and beam halo
muons with a photon from bremsstrahlung.  
Figure~\ref{fig:cdfgammatime} shows distributions of the
corrected arrival time for each background, the data,
and for a sample signal.  All background normalizations were
determined by fitting data in regions of $t_c^{\gamma}$
outside of the signal region.  Signal acceptance was
estimated using Monte Carlo simulations.

\begin{figure}[tb]
 \begin{center}
  \unitlength1cm
  \begin{picture}(12.5,6)(0,0)
   \put(0.0,0.0){\psfig{file=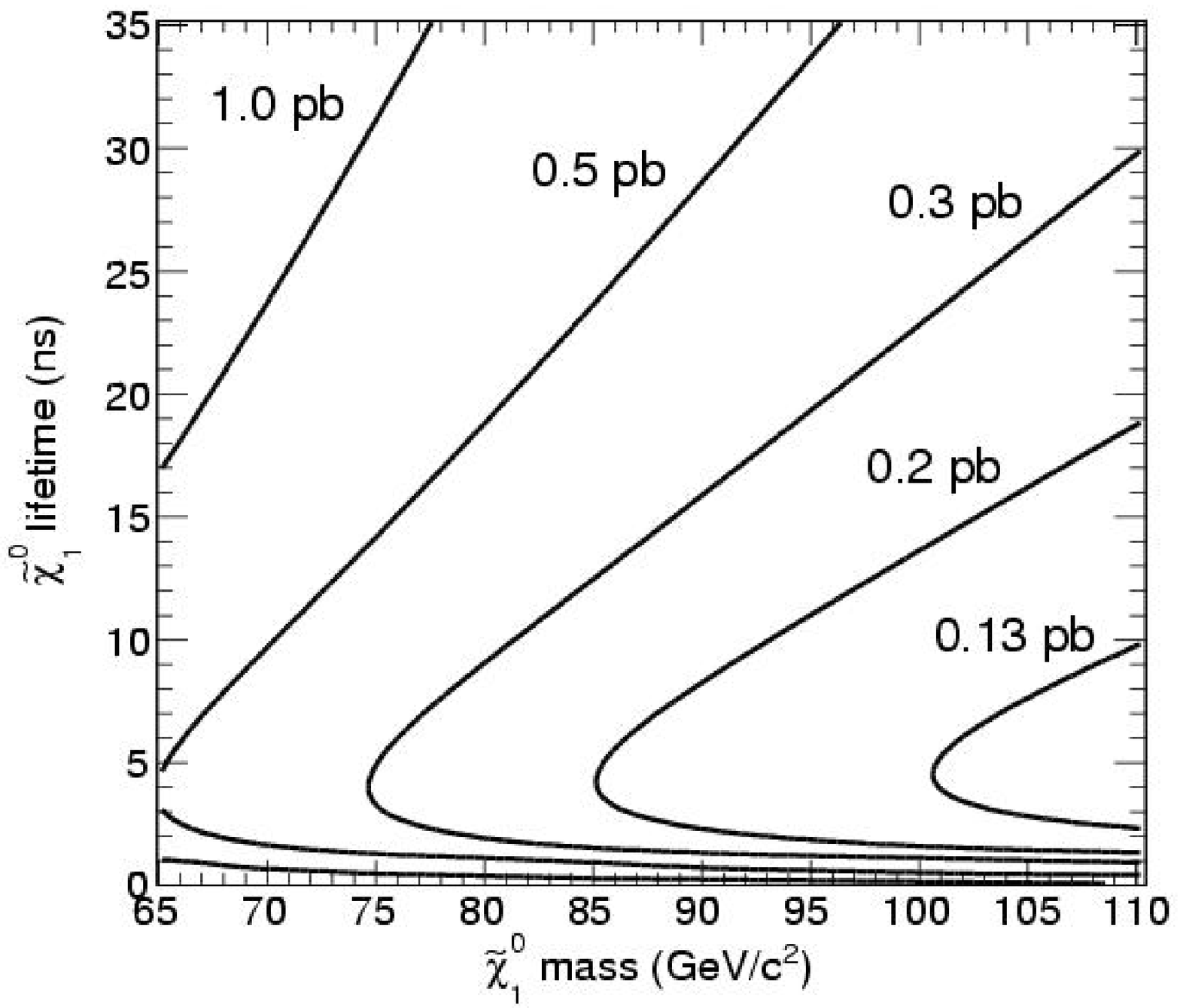,height=5.4cm}}
   \put(6.8,0.0){\psfig{file=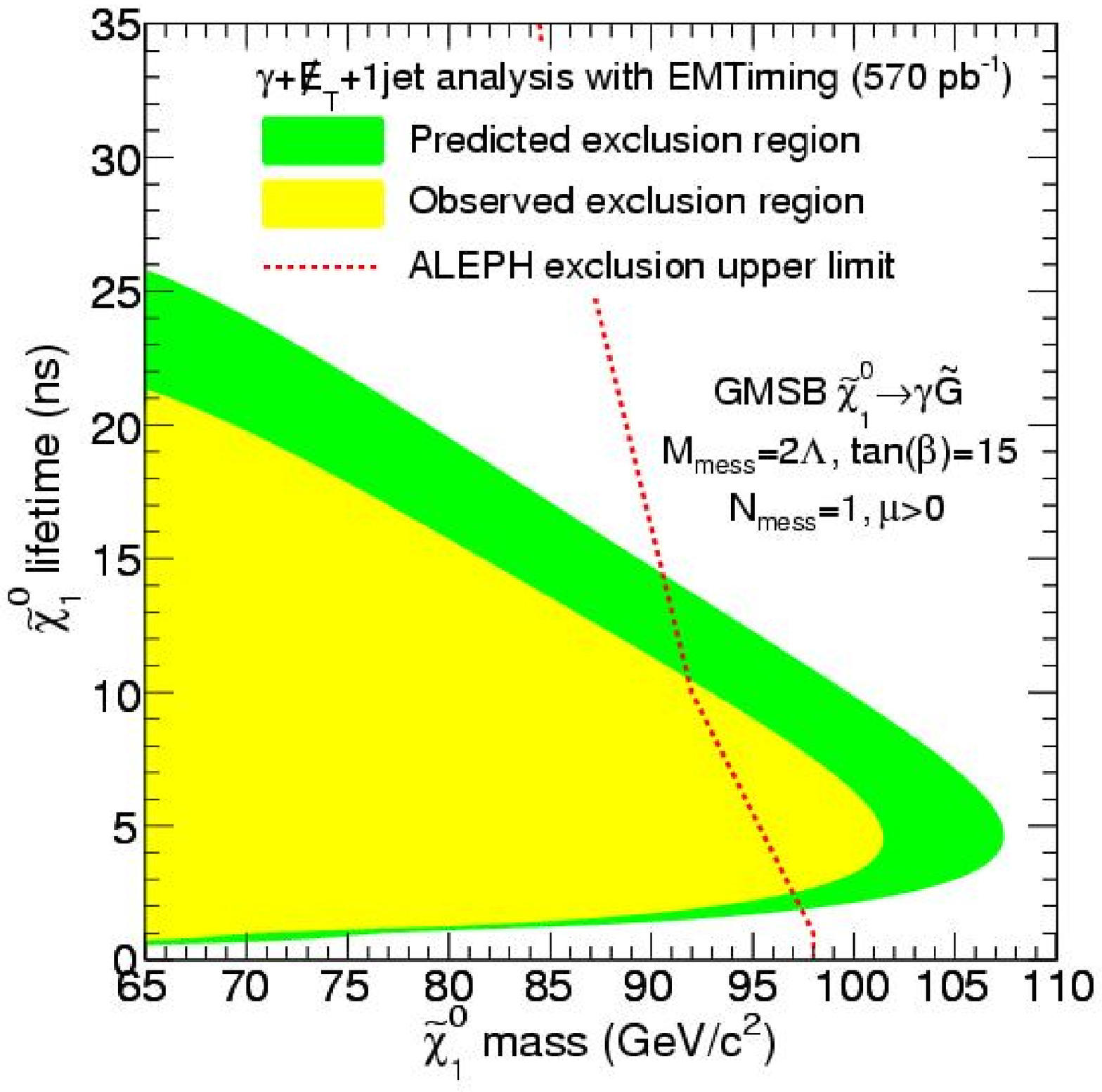,height=5.4cm}}
  \end{picture}
 \end{center}
\vspace*{8pt}
\caption{(left) 95\% CL cross section limit contours as a function
 of the \neutralino\ lifetime vs. mass.  (right) Expected and
 excluded regions of GMSB parameter space as a function of
 \neutralino\ lifetime vs. mass.
 Reprinted figures with permission from A.~Abulencia {\it et al.}  
 [CDF Collaboration], Phys.\ Rev.\ Lett.\ {\bf 99}, 121801 (2007).
 Copyright 2007 by the American Physical Society.
 \protect\label{fig:cdfgammalimits}}
\end{figure}

After optimization of several selection criteria,
background was estimated to be 1.25 $\pm$ 0.66 events
(0.71 $\pm$ 0.60 from SM, 0.46 $\pm$ 0.26 from cosmics,
0.07 $\pm$ 0.05 from beam halo).  Two events were observed
in data.  The signal acceptance depends on the mass
of the long-lived particle as well as its lifetime.  
Therefore, limits on the production cross section are
set in the lifetime versus mass plane (Fig.~\ref{fig:cdfgammalimits}).  
For the specific
GMSB model used to model the signal, the limits set
by CDF in the high mass region are greater than those from 
a previous analysis by ALEPH (see Ref.~\refcite{bib:aleph}).

\section{Stopped Particles \label{sec:stoppedgluinos}}	

D0 has carried out a search for very long-lived, heavy 
particles which stop within the calorimeter and decay 
microseconds to days later.\cite{bib:haas}  As
mentioned in Sec.~\ref{sec:theoryintro}, split
supersymmetry provides a plausible candidate.  In one 
scenario, charged R-hadrons lose energy via ionization and
some fraction can stop within the calorimeter (called
stopped gluinos).  At some later time the gluino can
decay to a gluon and a neutralino ($G \rightarrow g\neutralino$)
or a quark-antiquark and a neutralino 
($G \rightarrow q\bar{q}\neutralino$)
giving a large energy deposition in the calorimeter that is 
not associated (in time) with the interaction that produced it.

For such events, one expects a large energy deposit in
the calorimeter associated with large missing transverse energy and
no other significant event activity.
However, because of trigger limitations, the analysis is only sensitive
to gluino decays that occur during (or just before) a
bunch crossing.  Events were therefore selected that have no
activity in luminosity counters (located near the beamline)
indicative of no $p\bar{p}$ interaction.

\begin{figure}[tb]
 \begin{center}
  \unitlength1cm
  \begin{picture}(12.5,4)(0,0)
   \put(0.0,0.0){\psfig{file=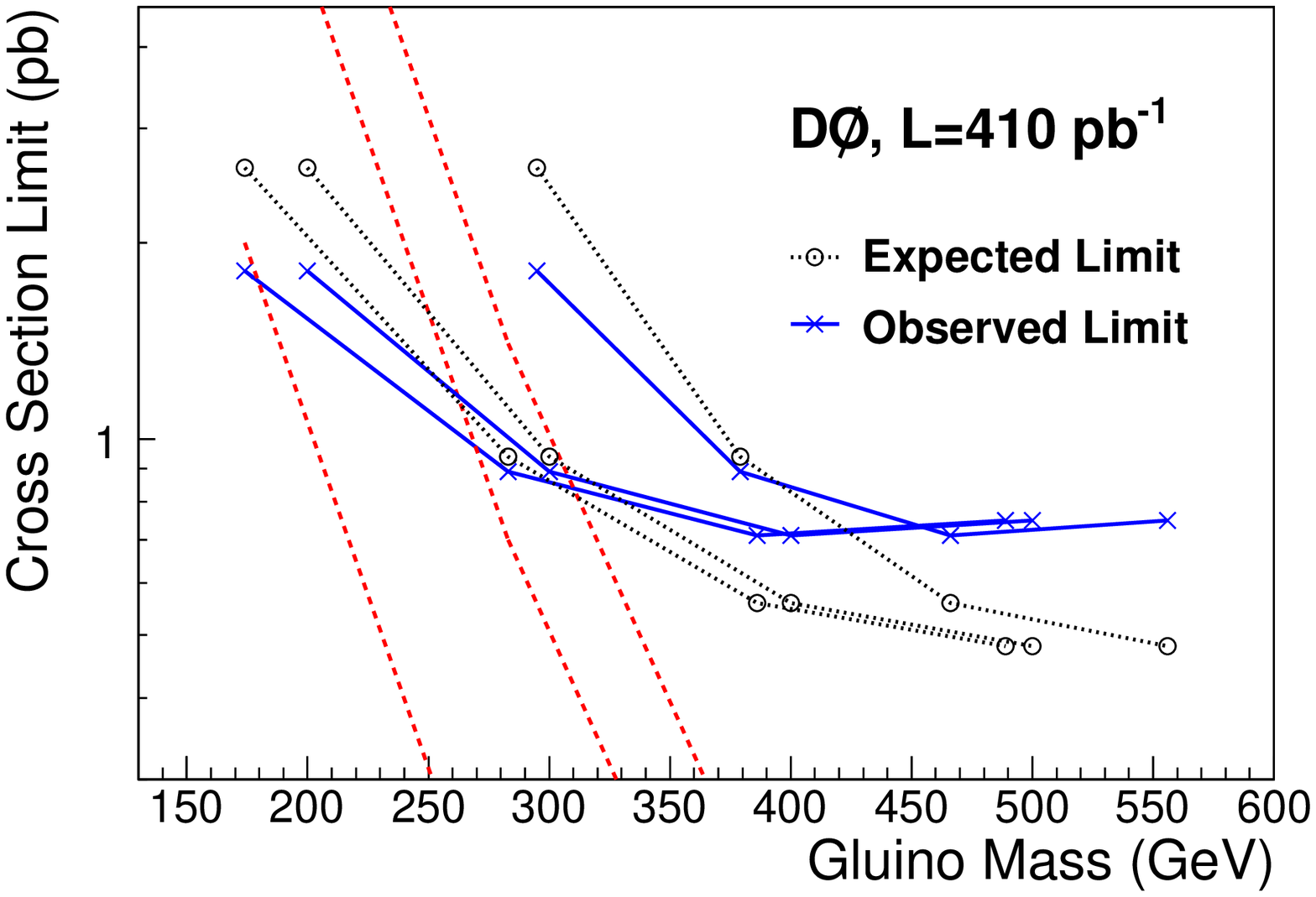,width=6cm}}
   \put(6.25,0.0){\psfig{file=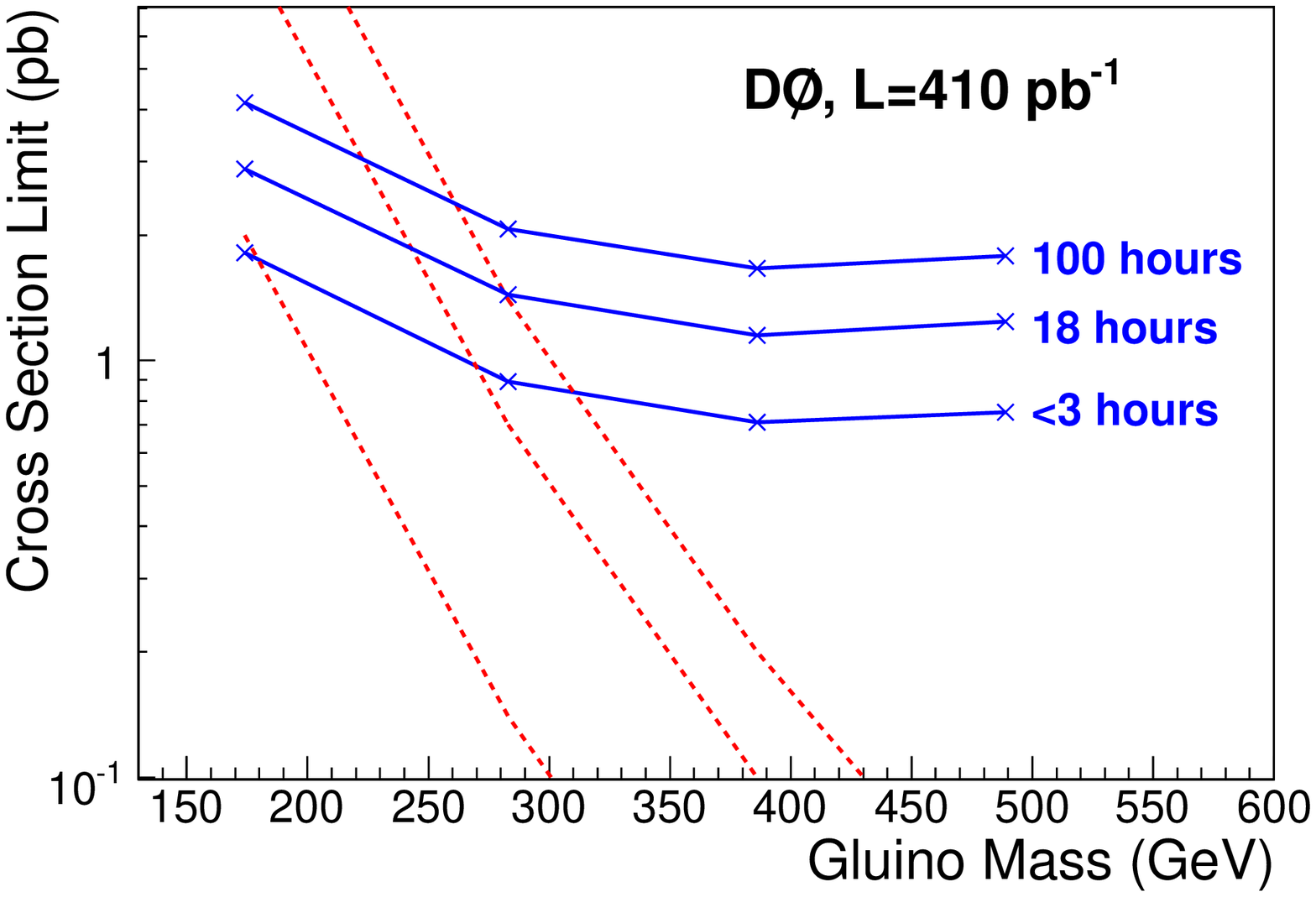,width=6cm}}
  \end{picture}
 \end{center}
\vspace*{8pt}
\caption{Cross section limits on stopped gluinos.  (left) Limits
assuming a lifetime $<$3 hours, 100\% BR to $g$\neutralino, and
three different \neutralino\ masses (50, 90, and 200 GeV, left to
right). The expected limit is shown in black and the observed limit
in blue.  The red lines indicate theoretical cross sections for
a range of assumed conversion cross sections.  (right) Limits
for a 50 GeV \neutralino\ mass for three different gluino
lifetimes ($<$3 hours, 18 hours, and 100 hours).
 Reprinted figures with permission from V.~M.~Abazov {\it et al.}  
 [D0 Collaboration], Phys.\ Rev.\ Lett.\  {\bf 99}, 131801 (2007).
 Copyright 2007 by the American Physical Society.
\protect\label{fig:d0stoppedgluinos}}
\end{figure}

Backgrounds are dominated by fake signals from cosmic muons 
and beam halo muons.  The selection criteria have reduced
the backgrounds from $p\bar{p}$ interactions to a negligible
level.  Fakes from cosmic and halo muons were estimated 
from complimentary data samples.  Multiple searches were
performed in bins of energy deposition in the calorimeter.  
It was found that in all cases the data
were consistent with estimated backgrounds.  The signal
sensitivity depends on the gluino lifetime and on the
mass of the LSP (\neutralino).  
Figure~\ref{fig:d0stoppedgluinos} shows the cross section
limit for gluino production as a function of gluino 
mass for different \neutralino\ masses and gluino 
lifetimes.  These limits are generally model-independent
and apply to any long-lived monojet signal.

\section{Charged Massive Stable Particles}	

The searches discussed above focused on neutral particles that 
decay to stable particles within the detector.  However,
new long-lived, charged particles are also possible.  If
the lifetime is short enough to decay within the central
tracking system, the most likely signature is a
charged track which has a kink (or additional tracks)
in the middle of it.  These topologies have been searched
for at LEP.\cite{bib:lepkinksearches}  New stable, 
charged particles are limited by cosmological 
constraints to masses below 230 GeV.\cite{bib:stablecharged}
In addition, the observation and measurement of the
mass/lifetime of a long-lived, charged particle 
could have implications on understanding of Big Bang 
nucleosynthesis.\cite{bib:takayama}

However, if the particle lives long enough to pass through 
the detector, it can leave other striking signatures.  A
massive particle will be slower moving than the typical
stable SM particle and will leave behind a greater
ionization trail (due to the effect of the particle mass in the
Bethe-Bloch formula).  If it does not interact strongly,
the particle will look like a slow moving, heavily ionizing 
muon.\cite{bib:drees}
Both CDF and D0 have analyses underway (with preliminary
results available and discussed below) that use timing 
techniques to look
for such particles.  A previous Run I analysis by 
CDF used a $dE/dx$ technique.\cite{bib:cdfrunIchamps}
A review of the theoretical issues and experiment
status of searches for massive, stable particles
is available in Ref.~\refcite{bib:smpreview}.

\subsection{CDF CHAMPs Search}

CDF has performed a search for CHArged Massive Particles
(CHAMPs) using the timing capabilities of its outer
tracker.\cite{bib:cdfchamps}  
Events are selected with at least one muon
(with $p_T >$ 20 GeV) that fired a single muon trigger
and is from the primary vertex.  Cosmic ray muons are rejected by 
looking for a second track where the pair is consistent
with one track moving inward and the other moving outward.
CHAMP candidate tracks include the trigger muon and
either a second muon or the leading non-muon.  Two
samples are used: the signal sample where both
candidates have $p_T >$ 40 GeV and a control sample
where both have $20 < p_T < 40$.

\begin{figure}[tb]
 \begin{center}
   {\psfig{file=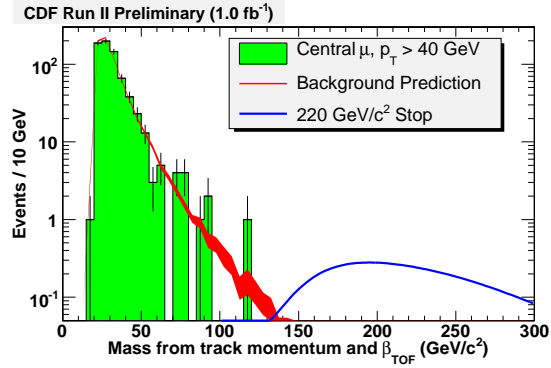,height=5cm}}
 \end{center}
\vspace*{8pt}
\caption{Mass distribution for signal events calculated using the
 TOF detector.  The filled histogram shows data while the red
 line indicates the expectation from the control region.  The
 blue line indicates the expectation for a 220 GeV stop mass.
 \protect\label{fig:cdfchampsmass}}
\end{figure}

For each CHAMP candidate, its velocity was measured from
its path length and the time it arrives at the 
time-of-flight (TOF) detector (part of the tracking
system).  Muons have a velocity $\beta = v/c = 1$ while
massive particles have $\beta < 1$.
The momentum and velocity were used to calculate the
candidate mass.  $W \rightarrow e\nu$ events were used
to study the efficiencies and the ability to model
backgrounds in the signal sample using the control
sample.  Figure~\ref{fig:cdfchampsmass} shows the
observed and predicted mass distribution for the
signal sample.  The region $M > 100$ GeV is
the final CHAMP sample.

\begin{figure}[tb]
 \begin{center}
   {\psfig{file=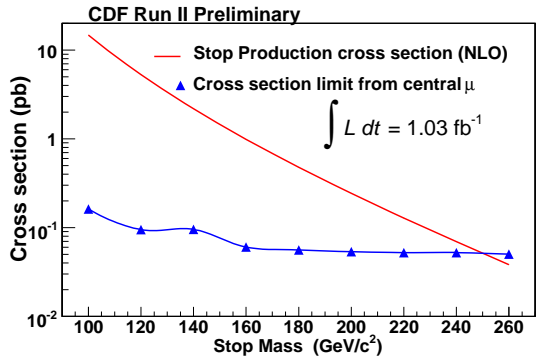,height=5cm}}
 \end{center}
\vspace*{8pt}
\caption{Observed limit on stop production cross section (blue line
 with points) as a function of mass.  The red line shows the 
 NLO production cross section.
 \protect\label{fig:cdfchampslimits}}
\end{figure}

The sample model used to estimate acceptance involved
pair production of long-lived stop squarks.\cite{bib:stopmodel}  
The stop hadronizes (similar to the gluinos described in
Sec.~\ref{sec:stoppedgluinos}) and hadronic effects
such as charge exchange were taken into account.  Only
one candidate CHAMP was found with $M > 100$ GeV,
consistent with the predicted background.  
Figure~\ref{fig:cdfchampslimits} shows the CDF preliminary 
limit on
the stable, massive, stop production cross section.
From the calculated next-to-leading order (NLO) cross 
section, a limit on the stop mass $>$240 GeV was set.

\subsection{D0 CMSP Search}

D0 labels its search as CMSP - charged, massive, stable
particles.\cite{bib:d0cmsp}  
It uses two different signal models as
possible sources of CMSPs.  The first is a GMSB
model with the $\tilde{\tau}$ as the NLSP.  Stau pair
production is modeled with a lifetime sufficiently large 
that both \stau 's decay outside the 
detector.\cite{bib:feng} The second model is an
anomaly-mediated supersymmetry breaking (AMSB) inspired
model where the LSP (\neutralino) and NLSP (\chargino) 
mass difference is sufficiently small ($<$150 MeV) to provide
a long lifetime to the \chargino.\cite{bib:masssplitting}
Two chargino cases are considered: (1) where the
chargino is mostly higgsino and (2) where it is
mostly wino.  

The analysis searches for pair production of 
long-lived staus or charginos having 
two CMSPs per event.  Therefore, events are
selected that contain two muons, which are generally
back-to-back in $\phi$.  For each muon, the 
average speed was calculated using the timing
information from each of the three layers of the muon 
scintillator system.  A muon travels at the speed
of light while a massive particle will move slower.
The sensitivity was therefore limited at lower
masses by the timing resolution and at higher
masses by the trigger that only accepted muons
within its timing window.

\begin{figure}[tb]
 \begin{center}
  \unitlength1cm
  \begin{picture}(12.5,4)(0,0)
   \put(0.0,0.0){\psfig{file=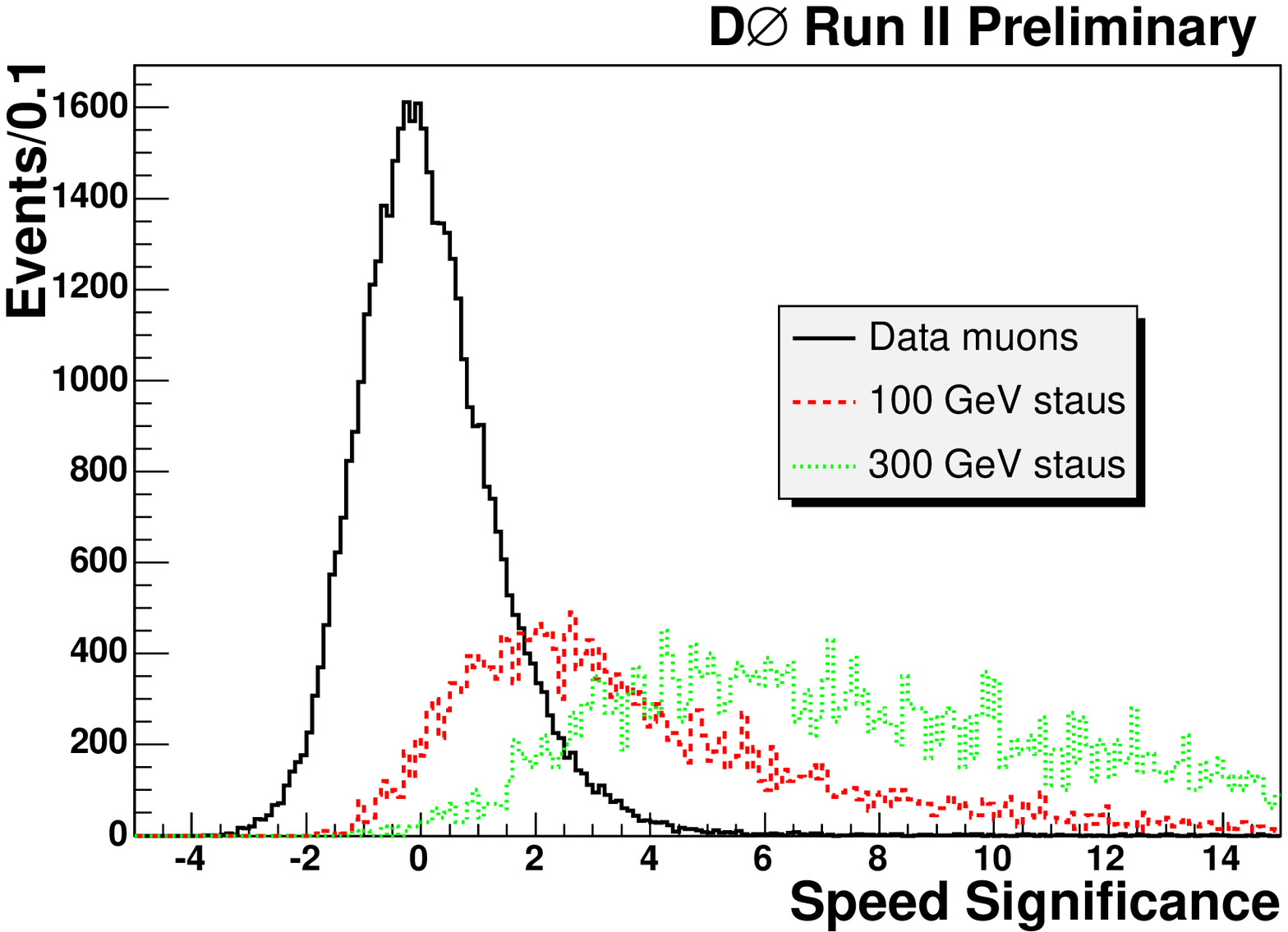,width=6cm}}
   \put(6.25,0.0){\psfig{file=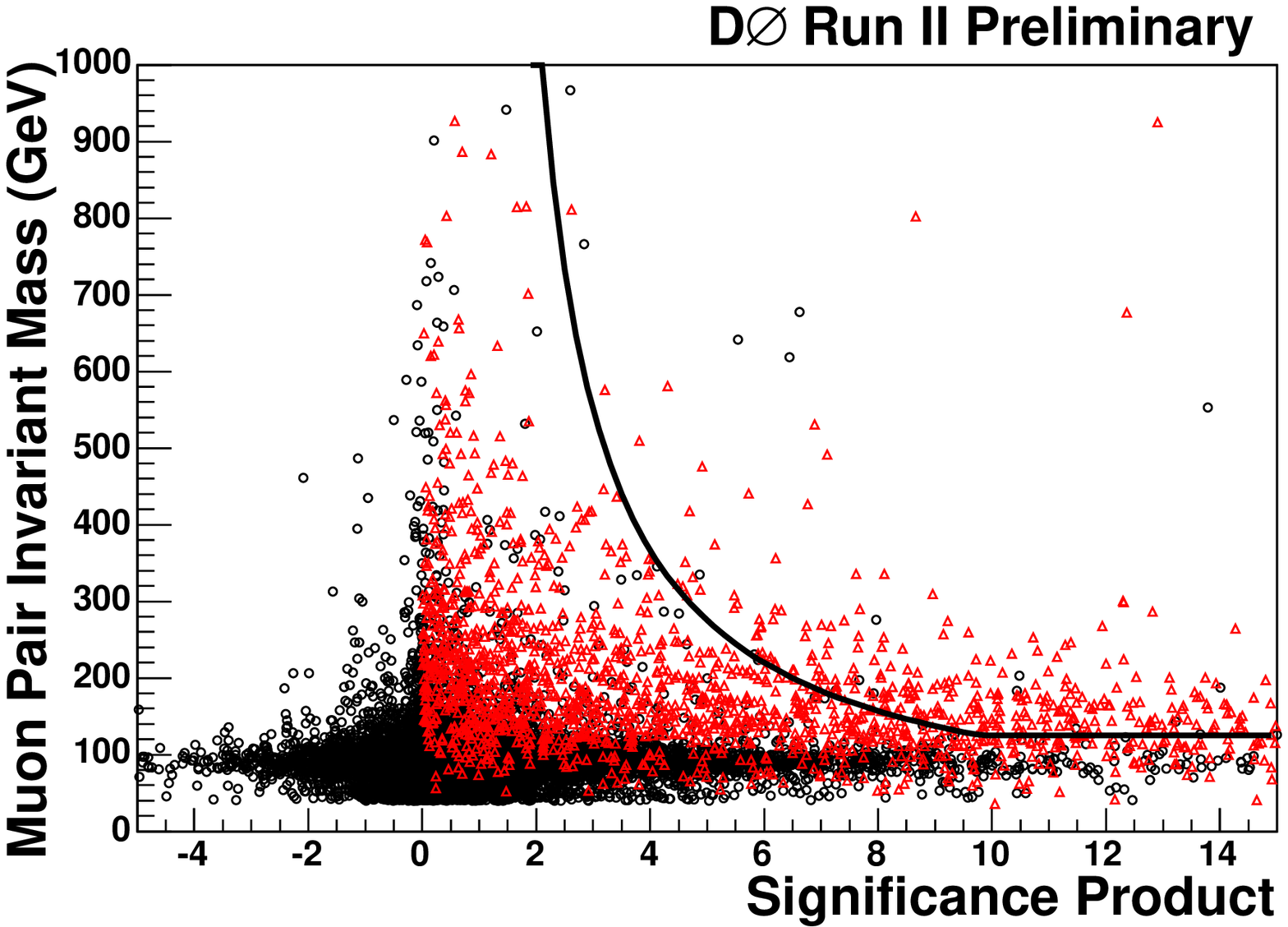,width=6cm}}
  \end{picture}
 \end{center}
\vspace*{8pt}
\caption{(left) The speed significance for data (black) and two
different stau masses (red and green).  (right) Distribution of
the invariant mass of the muon pair versus the speed 
significance product.  The black dots indicate data,
the red triangles show a 60 GeV stau,
while the black line shows a sample final analysis cut in
this parameter space.
\protect\label{fig:d0cmspvariables}}
\end{figure}

The average speed of the particle is used to calculate the 
speed significance:
\begin{equation}
  Significance = \frac{1 - speed}{\sigma_{speed}},
\end{equation}
where $speed$ is measured with respect to the speed
of light and $\sigma_{speed}$ is the uncertainty on
the speed measurement due to timing resolution.  Muons will have
a significance centered at zero while massive particles
are expected to have a large positive value 
(Fig.~\ref{fig:d0cmspvariables}(left)).  Events are required to
have both muons with positive speed significance.
The final analysis cut used the speed significance
product (the product of the two individual significances)
and the invariant mass of the two muons.  By cutting
in this two dimensional space, background was 
greatly reduced while retaining signal 
events (Fig.~\ref{fig:d0cmspvariables}(right)).  
This final criterion was optimized for six 
\stau\ signal masses: 60, 100, 150, 200, 250, and 300 GeV.

\begin{figure}[tbhp]
 \begin{center}
  \unitlength1cm
  \begin{picture}(12.5,4)(0,0)
   \put(0.0,0.0){\psfig{file=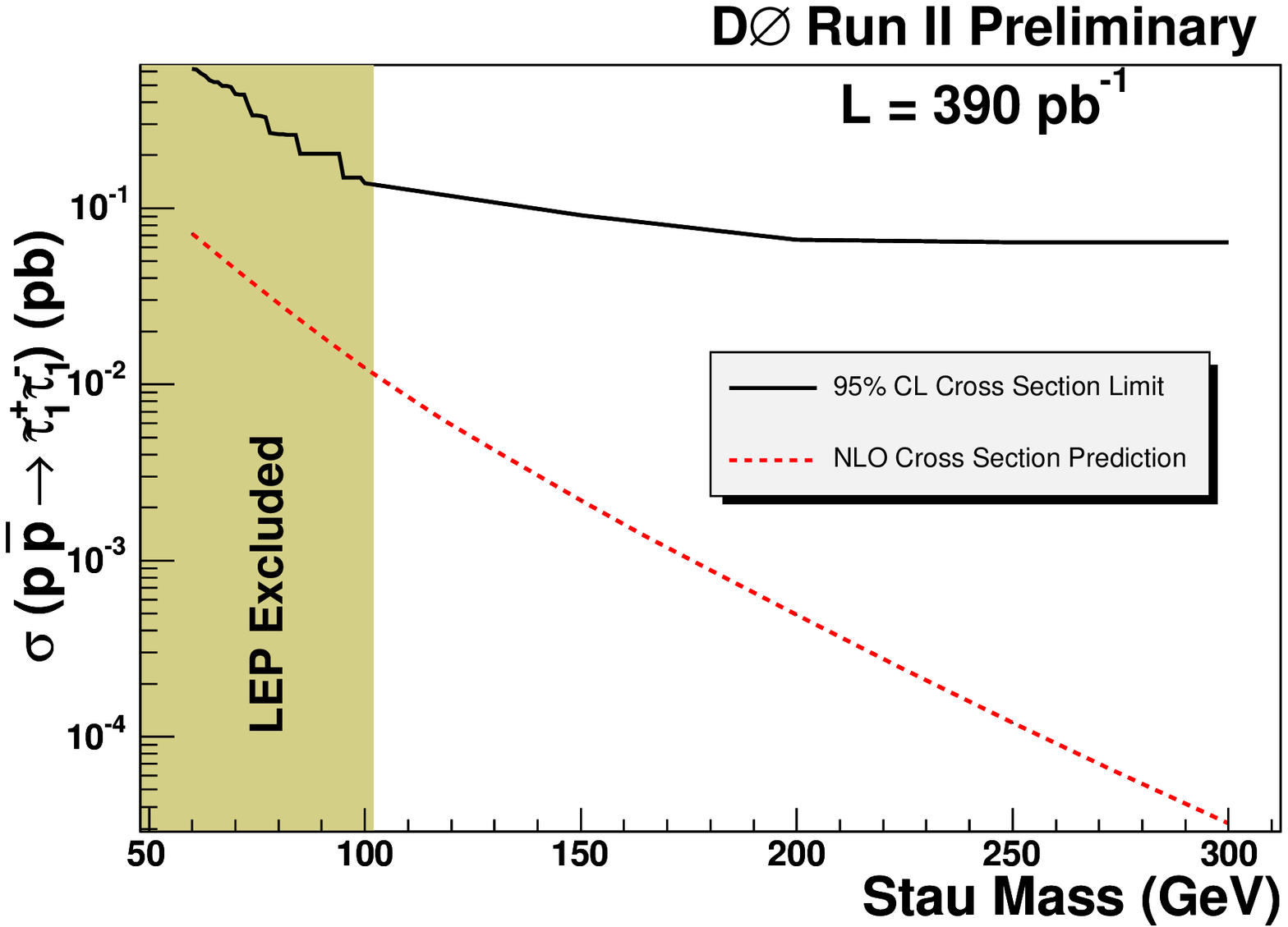,width=6cm}}
   \put(6.25,0.0){\psfig{file=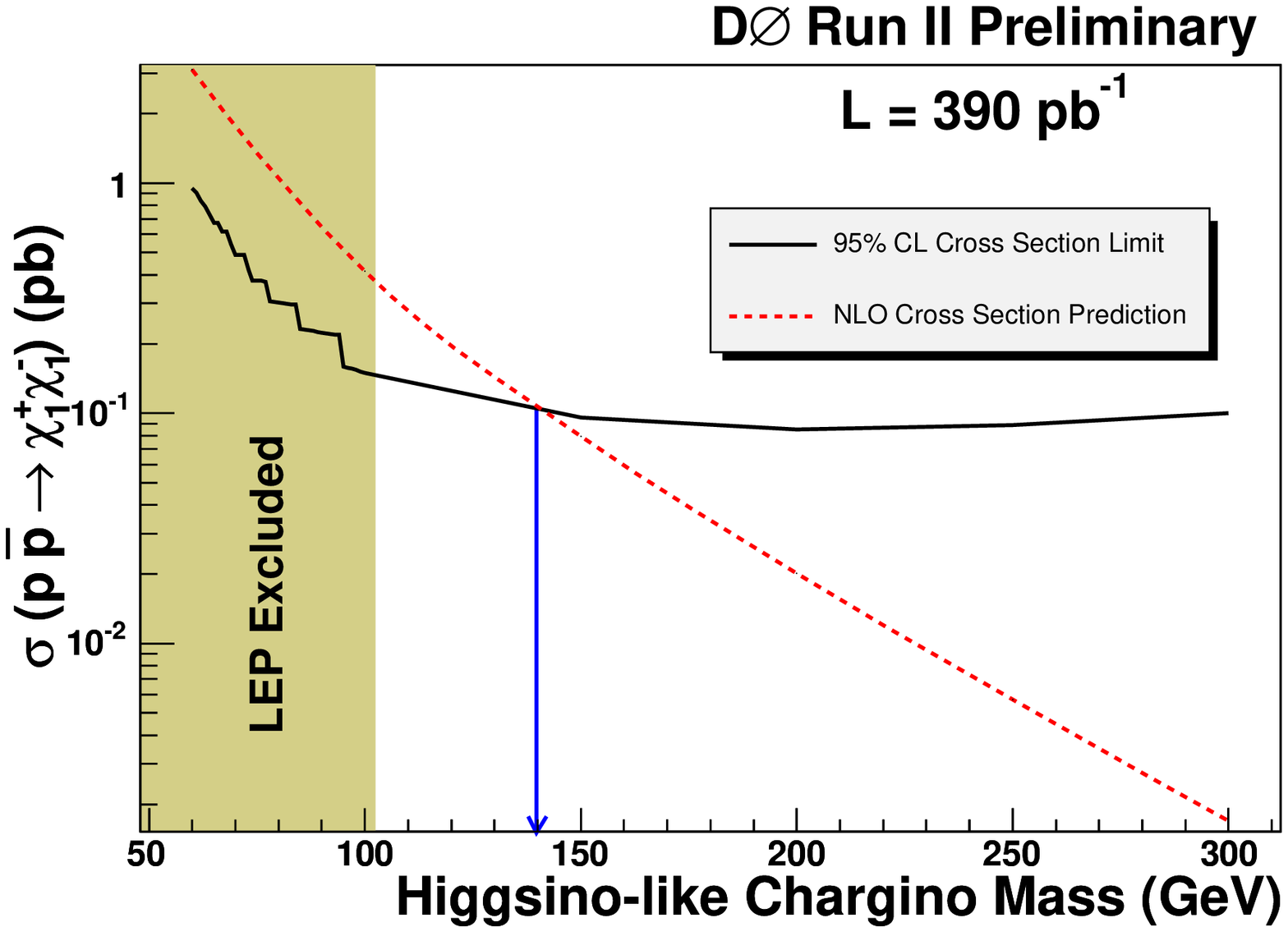,width=6cm}}
  \end{picture}
 \end{center}
\vspace*{8pt}
\caption{Limits on pair production of GMSB stau's and
AMSB chargino's from the D0 CMSP analysis.
\protect\label{fig:d0cmsplimits}}
\end{figure}

For all six masses, the observed data were consistent with
background, therefore preliminary limits were set on the \stau\ pair
production cross section (Fig.~\ref{fig:d0cmsplimits}(left)).  
However, the analysis is not yet sensitive to exclude masses beyond
the LEP limit.  Because the kinematics for chargino 
production are similar, the same cuts were applied and 
preliminary limits set on the chargino pair production cross 
section (Fig.~\ref{fig:d0cmsplimits}(right)).  Here, D0
excluded long-lived charginos with a mass below 174 GeV.
However, a shorter lifetime (such that some fraction of
the charginos decay before reaching the muon system) will
lower this limit. 

\section{Other Analyses Under Development}	

Several other analyses are underway and/or have been 
presented as preliminary searches for long-lived particles
by the CDF and D0 collaborations.  Brief descriptions 
of some are presented here.

CDF has searched for a neutral, massive, long-lived 
particle that decays to a $Z^0+X$ final state, such as
a new fourth generation quark 
$b^\prime \rightarrow b Z^0$.\cite{bib:cdfzparents}
The analysis searches for highly displaced vertices
($>$0.3 mm) with dimuon pairs whose invariant mass
falls within the $Z^0$ mass peak window 
(81 $< M <$ 101 GeV).  The acceptance is limited
by the trigger efficiency for muons with large DCA.
Separate searches are done
with and without a cut of $p_T >$ 30 GeV on the
$Z^0$ boson.  Both searches observe data 
consistent with background and limits are set on
the production cross section both as a function of
mass and lifetime.  

\begin{figure}[tbph]
 \begin{center}
   {\psfig{file=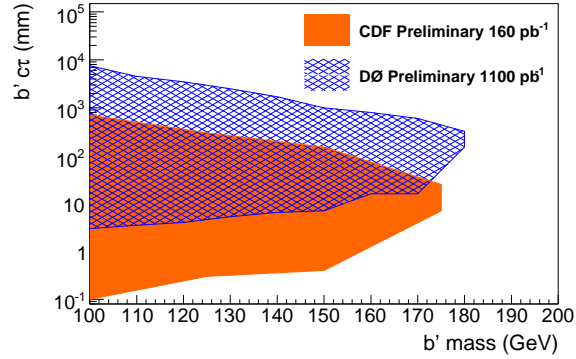,height=5cm}}
 \end{center}
\vspace*{8pt}
\caption{Limits on the $b^\prime$ model from CDF (red)
 and D0 (blue).
 \protect\label{fig:zparents}}
\end{figure}

D0 has developed a new technique to search for a similar
signature ($Z^0+X$) in the $Z^0 \rightarrow e^+e^-$
decay mode.\cite{bib:d0zparents}  By using the location
of hits in the central preshower (CPS) and four layers 
of the electromagnetic part of the calorimeter, the 
incident direction of the EM object can be measured.  
By not requiring a reconstructed track, the analysis
is sensitive to much longer lived parents.
Two EM objects are used to determine a vertex position
in the $x,y$ plane.  The radius is measured as
the distance from the PV to the EM vertex.  The sign
is determined by the sign of the cross product of the
radius vector and the $Z^0$ $p_T$ vector.  Negative
radii are used to model the background for 
events with positive
radii.  For events with an invariant mass $>$75 GeV, 
data are consistent with the estimated background.
Limits on the cross section are set.  While the
limits of both CDF and D0 are fairly model independent,
Fig.~\ref{fig:zparents} shows the preliminary limits on a 
long-lived $b^\prime$ model from both experiments.

CDF has searched for a long-lived, doubly charged 
Higgs boson by looking for heavily ionizing 
tracks.\cite{bib:cdfhiggspp}
$H^{\pm\pm}$ tracks will produce a signal several
times larger than for a minimum ionizing particle
in the central outer tracker (COT).  One search
looked for such signals while a complementary
search also required large energy deposition in
the calorimeter.  In both searches, the expected
background was $<10^{-4}$ events and zero events were
observed in data.  In the case of $H_L^{\pm\pm}$ and 
$H_R^{\pm\pm}$ degenerate in mass, a limit of
$H^{\pm\pm} < 146$ GeV was set.

The full capabilities of the CDF and D0 
experiments have yet to be realized.
The differences between the detectors have allowed
for a variety of search techniques looking
for long-lived particles.
Other Tevatron searches are possible including
kinked tracks and detached vertices using jets.
Combined with the larger data samples already recorded
by the two experiments, additional results can be
expected within the coming years.

\section{Summary}	

The CDF and D0 collaborations have performed a wide ranging
program of searches for long-lived particles that might
arise from new physics.  
Table~\ref{tab:summary} summarizes the types of analyses
and size of data samples used for each analysis.
These searches
greatly expand the discovery potential of the Tevatron
and provide experience that will be useful at the
LHC for similar work.
While new limits have been set within many BSM models 
(as discussed above), the greatest advantage of these 
analyses is the ability to explore new areas that may
be missed by traditional searches.  The lifetimes
explored cover many orders of magnitude.

In addition to the new data being accumulated at the 
Tevatron, new techniques are being developed to continue
the exploration of additional possibilities.  The indirect
evidence for physics beyond the SM is overwhelming, but
the lack of direct evidence drives our need to look 
simultaneously in many new directions in the hope that
at least one will be fruitful.  The Fermilab program is doing an
excellent job in this regard.

\begin{table}
 \tbl{Summary of Tevatron searches for long-lived particles.
      \label{tab:summary}}
  {\begin{tabular}{lccc}
    & & Luminosity & \\ \toprule
    Signal & Collaboration & (pb$^{-1}$) & Limits \\ \toprule
    NLLP decaying to muon pairs    & D0  &  380 & NuTeV excess \\
    NLLP decaying to photon + MET  & CDF &  570 & GMSB SUSY \\
    Stopped gluinos                & D0  &  410 & split supersymmetry \\
    CHAMPS                         & CDF & 1000 & stop \\
    CMSP                           & D0  &  390 & \stau, \neutralino \\
    LL $Z^0$ parents               & CDF &  163 & $b^\prime$ \\
    LL $Z^0$ parents               & D0  & 1100 & $b^\prime$ \\
    LL $H^{\pm\pm}$                & CDF &  292 & charged Higgs \\
     \botrule
  \end{tabular}}
\end{table}

\section*{Acknowledgments}

The author wishes to thank the CDF and D0 collaborations for producing,
publishing and making available the information and plots
used throughout this review.
We'd like to thank Jill Adams for careful reading
and Harrison Prosper for his insightful comments.  
This work is supported in part by the U.S. Department of Energy.

\end{document}